\documentclass[]{tMOP2e}
\usepackage{amsmath}
\usepackage{graphicx}
\usepackage{amssymb}

\begin{document}
\newcommand {\rmp} {Rev. Mod. Phys.}
\newcommand {\prb} {Phys. Rev. B.}
\newcommand {\prl} {Rev. Mod. Lett.}

\title{Spectroscopic Characterization of Gapped Graphene in
the Presence of Circularly Polarized Light }
\author{ Godfrey Gumbs$^1$, O. Roslyak$^1$, Danhong Huang$^3$, Antonios Balassis$^2$ 
\\\vspace{6pt} 
$^1${ \em {Department of Physics and Astronomy,
Hunter College at the City University of New York,
695 Park Avenue New York, NY 10065, USA.}}\\
$^2${ \em {Physics Department, Fordham University, 441 East Fordham Road,
       Bronx, NY 10458, USA.}}\\
$^3${\em{Air Force Research Laboratory, Space Vehicles
Directorate,\\
Kirtland Air Force Base, NM 87117, USA.}}
}

\date{\today}

\maketitle

\begin{abstract}
We present a description of the energy loss  of a charged particle
moving parallel to  a graphene layer and graphene double layers. Specifically,
we compare the stopping power of the plasma oscillations for these two configurations
in the absence as well as the presence of circularly polarized light whose frequency
and intensity can be varied to yield an energy  gap of several hundred $\texttt{meV}$
between the valence and conduction  bands. The dressed states of the Dirac electrons
by the photons yield collective plasma excitations  whose characteristics are qualitatively
and quantitatively different from those produced by Dirac fermions in gapless graphene,
due in  part to the finite effective mass of the dressed electrons. For example,
the range of wave numbers for undamped self-sustaining plasmons is increased
as the gap is increased, thereby increasing the stopping
power of graphene for some range of charged particle velocity when graphene
is radiated by circularly polarized light.
\end{abstract}

\section{Introduction}
\label{sec1}

Graphene~\cite{wallace} has now been investigated intensively both theoretically
and experimentally to gain a clearer understanding of its properties
to see how it can be employed in nanoelectronic device applications. 
Several review
articles have drawn together many aspects of graphene with regard to the role which
many-body effects have on their electron transport and optical response
properties~\cite{neto,chakraborty,philtrans}. 
However, we still need to
explore in greater detail the energy loss when a beam of electrons
travels near a single layer or double layers of graphene. 
Generally,
inelastic interactions include phonon excitations, inter and intra-band
transitions, plasmon excitations, inner shell ionizations, and Cerenkov
radiation. 
This is a non-trivial challenge which we investigate by considering
a model which accounts for only plasma losses and those due to intra-band and
inter-band transitions. 
The beam of incident electrons is assumed to be focused in a narrow angle
with a known, small range of kinetic energies. 
Some initial work has been reported
in Refs.~\cite{Miscovic-1,Miscovic-2,fessatidis} for gapless graphene. 
In this manuscript, our interest is to investigate how the stopping power is affected when a
gap opens up between the valence and conduction bands, as may occur, for instance, 
when graphene is radiated by a circularly polarized electric field (CPEF).

 \par
Electron energy loss spectroscopy (EELS) has been considered for many systems dating
back to Ritchie's classic paper\cite{ritchie} for a slab of dielectric material in the
local limit and subsequently generalized to a non-local dielectric function by Gumbs
and Horing~\cite{slab}. The nonlocality was included in Ref.~\cite{slab} through
the screening of the electron-electron interaction produced by single-particle excitations and
plasmon modes. Fessatidis, et al.~\cite{fessatidis} recently adopted the formalism of
Horing, et al.\cite{tso} for a 2D electron gas and Gumbs and Balassis~\cite{GG} for a
nanotube to the case of gapless graphene with the aid of the polarization function calculated by
Wunsch, et al.~\cite{wunsch} for conventional Dirac electrons.

\par
There have been several recent papers which considered  the effects which a
CPEF~\cite{kibis,roslyak}, spin-orbit interaction (SOI)~\cite{SOI} in suspended
graphene or the sublattice symmetry breaking (SSB)
by an underlying polar substrate~\cite{SOI-2,li_2009,giovannetti_2007} in epitaxial
graphene may have on the energy band structure and plasma excitations~\cite{roslyak} of
a graphene sheet. 
Under these conditions, there is a gap  between the valence and
conduction bands as well as between the intra-band and inter-band electron-hole
continuum of the otherwise semimetal Dirac system.~\cite{wunsch,shung1,shung2}
Additionally, the interplay between the single-particle excitations in the long wavelength
limit results in dielectric screening of the Coulomb interaction which produces
an undamped plasmon mode which appears in the gap separating the two
types of electron-hole modes forming a continuum. 
By this, we mean that a
self-sustained collective plasma mode is not supported by exciting \emph{either}
the intra-band or inter-band single-particle modes only. 
Furthermore,
although the Dirac electrons near the $\mathbf{K}$-points now acquire a non-zero effective
mass with a CPEF, one still cannot produce a long wavelength plasmon mode
as it is possible in the two-dimensional electron gas (2DEG) by intra-band excitations only. 
These properties of the plasma modes give rise to noticeable differences in the behavior
of the stopping power of gapped graphene compared with conventional graphene.

\par
A  dynamic energy gap produced by the CPEF may be comparable with that
induced by a polar substrate. However, unlike the latter, the gap produced
by a CPEF is highly tunable. 
Additionally, those two gap-opening mechanisms
may take effect simultaneously. 
Whenever this occurs, the circular polarization
of the electromagnetic  field may no longer be required provided that it is
in resonance with the SSB gap. 
Electron energy loss spectroscopy (EELS) may
be employed to ascertain the plasmon frequencies in single and double layer graphene.
This setup comprises an electron/photon mixed pump-probe, 
with the electron beam being the probe
effected by the polarized light whose role
is to open the gap. 
The Raman shift of the scattered electrons provides both particle-hole
and plasmon excitation frequencies, 
which are usually characterized by their spectral weight,
a quantity that  depends on the
transferred energy $\hbar \omega$ and momentum $\hbar q$.

\par
The outline of the  remainder of this paper is as follows. In Sec.\ \ref{sec2}, we present
the model Hamiltonian for Dirac electrons in the presence of a photon field. We present
the energy eigenvalues and eigenfunctions of the dressed Dirac electrons in the presence of circularly
polarized light.  We present the RPA  result for the polarization matrix for a double layer
in Sec.\ \ref{sec3} which we then employ  to obtain the energy loss for a
charged particle moving parallel to the  graphene   sheets.
 Our numerical results for the energy loss are presented in Sec.\ \ref{sec4}.

\section{Dressed Dirac electrons and Floquet sub-bands.}
\label{sec2}

Let us consider several graphene layers separated by distance $d$ from each other.
The layers may be epitaxially grown on a carbon or silicon based polar substrate.
In the latter case the system it to be embedded into a micro-cavity (MC) for effective electron dressing.
The central MC mode is characterized by associated energy $N_0 \hbar \omega_0$, 
with the mode being tuned with the substrate induced gap.
Here, $N_0$ is the average number of photons in the mode, and depends on specifics of the excitation method.
For example, if the mode is thermally excited then $N_0 \approx \hbar \omega_0/k_B T$ for small
thermal energy $k_B T$. 
Another method of excitation is to use an external circularly polarized
laser beam making $N_0$ a function of the external pump intensity.
This dressing method is suitable for gapless free standing graphene. 
In our formalism,
specific details of the method of excitation do not play a crucial role and will be neglected.
Additionally, we assume that all graphene layers are in the node of the optical filed and
thus we neglect retardation effects. This is valid if $d \ll 2 \pi/\omega_0$.

\par
In order to proceed further, we must specify the model
Hamiltonian at the two inequivalent $\mathbf{K}$ and $\mathbf{K}^\prime$
points in the graphene energy dispersion, i.e.,

\begin{gather}
\label{EQ:HAMILTONIAN}
\mathcal{H}=\mathcal{H}_{\rm JC}+\mathcal{H}_{\rm D}\\
\label{EQ:JAMESCUMMINSHAMILTONIAN}
\mathcal{H}_{\rm JC} = \hbar \omega_0 a^\dag a - \frac{W_0}{2 \sqrt{N_0}} \left({\sigma_{+} a + \sigma_{-} a^\dag}\right)\\
\label{EQ:DIRACHAMILTONIAN}
\mathcal{H}_{\rm D} = \hbar \it{v}_F \left({\left({\sigma_{+} + \sigma_{-}}\right) k_x \pm i\left({\sigma_{-} - \sigma_{+}}\right) k_y}\right)\ .
\end{gather}
The operators involved
in the above Hamiltonian act in the joint electron-photon space via the following
relations:

\begin{eqnarray*}
\sigma_{+} = \vert \uparrow \rangle \langle \downarrow \vert \; ; &
\sigma_{-} = \vert \downarrow \rangle \langle \uparrow \vert \; ; \\
\sigma_{+} \vert \uparrow, N_0 \rangle = 0 \; ; & \sigma_{+} \vert \downarrow, N_0 \rangle = \vert \uparrow, N_0 \rangle \\
\sigma_{-} \vert \downarrow, N_0 \rangle = 0\; ; & \sigma_{-} \vert \uparrow, N_0 \rangle = \vert \downarrow, N_0 \rangle \\
a \vert \downarrow \uparrow, N_0 \rangle = \sqrt{N_0} \vert \downarrow \uparrow, N_0-1 \rangle\;
; & a^\dag \vert \downarrow \uparrow, N_0 \rangle = \sqrt{N_0+1} \vert \downarrow \uparrow, N_0+1 \rangle
\end{eqnarray*}

with $\vert{\uparrow \downarrow}\rangle$ being the Dirac pseudo-spin basis.
In this notation, the Jaynes-Cummings (JC) Hamiltonian \eqref{EQ:JAMESCUMMINSHAMILTONIAN}
takes into account  the circularly polarized field of frequency $\omega_0$ and
 amplitude $\sqrt{4 \pi N_0 \hbar \omega_0/V}$. Also, $V$ is the characteristic volume of the mode. 
The Dirac part of the
Hamiltonian is given by Eq.~\eqref{EQ:DIRACHAMILTONIAN}, with $\pm$ corresponding
to the $\mathbf{K},\mathbf{K}^\prime$ points, respectively. 
\par
The energy of an electron rotational motion induced by the field is denoted by
$W_0 \propto \sqrt{N_0}$.
Provided it is assumed to be much less than the energy of the
optical filed itself, i.e.,  $W_0/\hbar \omega_0 \ll 1$, 
the Hamiltonian may be further simplified\cite{kibis} to:
$
H = \mathbb{I} N_0 \hbar \omega_0 + \left({E_g/2}\right) \sigma_3 +  \mathcal{H}_D
$
where $E_g = \sqrt{W^2_0 + \Delta^2}-\hbar \omega_0$ is the energy gap between valence
and conduction bands.
The gap provides dressed Dirac electrons with an effective mass $m^\star = E_g/\left({2 \hbar^2 v^2_F}\right)$.  
Corresponding eigenvalues and eigenfunctions summarized in Table~\ref{default}
\footnote{Note that the dressing does not mix the vales. 
Only $\mathbf{K}$ point is rpresented in the table since the relevant parameters (transition energies and the structure factor) are valey independent.}
. 
For the reader's convenience, dressed and free Dirac electrons are compared. 
It is a simple matter to show that the latter are just the off-resonance limit of the dressed states.

\par
There is an alternative approach to the problem of the gap opening by light  based on classical description of light\cite{OKA:2009,SYZRANOV:2008,ZHOU:2011,CALVO:2011}.
In their approach the quasi-energies of the optically induced sub-bands were obtained by exploiting the Floquet theorem (See Eq.(5) in Ref.\cite{ZHOU:2011}).
When the laser field of intensity $\sim A^2_0$ changes its polarization from being linear $(\phi =0)$ to
circular polarized $(\phi = pi/2)$ the degeneracy of the subbands $(\pm \hbar v_F k + N_0 \hbar \Omega ) $ is lifted  at the crossing points. 
The dynamic gap occurs at the Dirac point $(k=0)$.
It is proportional to the intensity of the optical field:
$E_g = 2 \sin(\phi)  (e A_0 v_F)^2/ \hbar \Omega$.
The dynamic gap also occurs at the cross-points away from the Dirac point $(k=\hbar \omega /2 v_f)$.
It is proportional to the amplitude of the optical field: $E_g = e A_0 v_F \sqrt{1- \cos(\phi) \sin(2 \alpha)}$,
where $\tan(\alpha) = k_y/k_x$.
\par
A few words must be said about the effective chemical potential for the Floquet states, which
separates the occupied (hole-like) from the unoccupied (electron-like) states in
quasi-equilibrium.
Its role is played by the "mean" energy (See Eq.(13) in Ref.\cite{ZHOU:2011}), 
which depends on the frequency of the light.
It will be used in the next section to define the distribution of the Floquet states (dressed Dirac electrons).
\par
In conventional semiconductors the optically induced side-bands are responsible for effects such as photon-assisted tunneling\cite{GRIFONI:1998,KOHLER:2005}, generation of excitons\cite{CERNE:1997,NORDSTROM:1998}, Stark\cite{NORDSTROM:1998} and dynamical Frantz-Keldish phenomena\cite{SRIVASTAVA:2004,ZHANG:2006}.  
In graphene, the side bands formation was analyzed in connection with photo-voltaic Hall effect\cite{OKA:2009}, dc transport\cite{CALVO:2011} and pump-probe optical response\cite{ZHOU:2011}.
The next section focuses on yet another effect  - electron energy loss via inelastic scattering by plasmons.

\section{ Energy Loss Formalism}
\label{sec3}

Let us consider a charged particle with charge $Ze$ moving parallel 
to the graphene surface with velocity
$\textbf{v}$ at a distance $z_0$ from the  top layer of a pair
of graphene sheets with separation $d$. Then, the rate of loss
of energy of this charged particle
due to the frictional force it experiences by passing over the
interacting electron gas is given by

\begin{equation}
\frac{dW}{dt}=-2\pi \frac{(Ze)^2}{4\pi\varepsilon_0}
 \int dz^{\prime} \int \frac{d^2\textbf{q}}{(2\pi)^2}
(i\textbf{q} \cdot \textbf{v})\frac{e^{-q|z^\prime - z_0|}}{q}
\Im \text{m} \
\epsilon^{-1}(z_0,z^\prime; q, -\textbf{q}\cdot \textbf{v})
\label{parallel}
\end{equation}
where $\epsilon^{-1}(z,z^\prime; \textbf{q},\omega)$ is the inverse
dielectric function. 
For a double layer, we may follow the procedure presented in 
Ref.~\cite{EELS_SSC} to obtain the inverse dielectric function as
a single layer may be described by the same formalism by setting the
inter-layer distance to infinity. Choosing the origin $z=0$ at one of the layers,
we obtain the inverse dielectric function as~\cite{EELS_SSC}

\begin{equation}
\epsilon^{-1}(z,z^\prime;q,\omega)= \delta(z-z^\prime)   +  \sum_{j,j^\prime=0,1}
\delta\left({z^\prime-(j-1)d}\right)  v_c(jd,j^\prime d)\Pi_{jj^\prime}(q,\omega)
\label{2-layers}
\end{equation}
where $\delta(z-z^\prime)$ is the Dirac delta function and polarization matrix is

\begin{equation}
\left({
\begin{array}{cc}
\Pi_{11} & \Pi_{12}\\
\Pi_{21} & \Pi_{22}
\end{array}
}\right)
=
\frac{1}{\epsilon(q,\omega)}
\left({
\begin{array}{cc}
\Pi^{(0)}_{11} \left({1-v_c(q)\Pi^{(0)}_{22} }\right) & v_c(q)\ e^{-qd}\Pi^{(0)}_{11} \Pi^{(0)}_{22}\\
v_c(q)\ e^{-qd}\Pi^{(0)}_{11} \Pi^{(0)}_{22} & \Pi^{(0)}_{22} \left({1-v_c(q)\Pi^{(0)}_{11}}\right)
\end{array}
}\right)
\label{EQ:POLARIZATIONMATRIX}
\end{equation}

It is expressed in terms of the generalized dielectric function

\begin{equation}
\label{EQ:DETERMINANTSINGLE}
\epsilon(q,\omega)= 1-v_c(q)\left\{\left[\Pi^{(0)}_{11} +\Pi^{(0)}_{22}  \right]+\left[
\left( 1 - e^{-2qd} \right)v_c(q)\Pi^{(0)}_{11} \Pi^{(0)}_{22}  \right] \right \} \ ,
\end{equation}
where $v_c(q) = 2 \pi e^2/\epsilon_s q$ with $\epsilon_s\equiv 4\pi\varepsilon_0\epsilon_b$,
 $\epsilon_b$ being the average background dielectric constant.
Noninteracting polarization of the given graphene layer assumes Lindhard form

\begin{equation}
 \Pi^{(0)}(q,\omega)_{11(22)} = \frac{1}{A}
\sum_{{\bf k},\ \lambda,\lambda^\prime=\pm1}
\frac{f_0(E_{{\bf k},\lambda})-f_0(E_{{\bf k}+{\bf q},\lambda^\prime})}
{\hbar\omega-E_{{\bf k}+{\bf q},\lambda^\prime}+E_{{\bf k},\lambda}}
F_{\textbf{k},\textbf{k}+\textbf{q}}^{\lambda\lambda^\prime} \ ,
\label{eps}
\end{equation}
Here, we have $E_{\mathbf{k},\lambda} =\lambda\hbar v_F  k$ for free Dirac
electrons, but
$E_{\mathbf{k},\lambda} =\hbar v_F\left\{ N_0 \omega_0 + \lambda \sqrt{(\epsilon_g/2)^2
+ k^2}\right\}$ for the dressed ones.
Also, the corresponding structure factors $\mathcal{F}^{\beta}_{\mathbf{k},\mathbf{k+q}}$ and
particle distributions $f_0(E_{{\bf k},\lambda})$ are given in Table~\ref{default}. 
$A$ is the normalization area in the reciprocal space chosen to assure $(1/A) \sum \limits_{\mathbf{k}}
 \approx  \int d^2 \mathbf{k}$. The upper limit of the integration is given by the ultraviolet cut-off as in Ref.~\cite{wunsch}.
Straightforward substitution of Eq. \eqref{2-layers} into \eqref{parallel} yield the
energy loss for graphene double-layer:
\begin{gather}
\label{EQ:STOPINGPOWER}
\frac{dW}{dt}  = \mathbf{F} \cdot \mathbf{v}_{\parallel}  = 2\pi  \ \frac{(Ze)^2}{4\pi\varepsilon_0}  \int \frac{d^2\textbf{q}}{(2\pi)^2}
\frac{(\textbf{q} \cdot \textbf{v} )}{q} v_c(q) \times \\ 
\notag
\times
\Im \text{m}\
\left\{  e^{-2q z_0}
\Pi_{11}(q, -\textbf{q}\cdot \textbf{v})
+ e^{-2q (z_0+d)}
\Pi_{22}(q, -\textbf{q}\cdot \textbf{v}) 
+ 2e^{-q(2 z_0+d)}
\Pi_{12}(q, -\textbf{q}\cdot \textbf{v})
\right\}
\end{gather}
The single layer stopping power may be obtained from 
Eqs.~(\ref{EQ:POLARIZATIONMATRIX},\ref{EQ:DETERMINANTSINGLE},\ref{EQ:STOPINGPOWER})
 by putting $\Pi^{(0)}_{22}(q,\omega) = 0$.
\par
The contributions to the polarization matrix and consequently to the rate of loss of energy in the above
Eq.~\eqref{EQ:STOPINGPOWER} may be identified as
coming from two terms \cite{tso,GG,fessatidis}.
One is determined by the plasmon excitations, i.e., when \emph{both} the real and imaginary parts of
$\epsilon(q,\omega)$ are simultaneously equal to zero.
The other comes from single-particle excitations, i.e., whenever
$\Im \text{m}\ \epsilon(q,\omega)$ is non-zero.  
Their relative contribution as a function of the energy gap induced by the dressing is
the subject of the next section.

\section{Numerical results and Discussion}
\label{sec4}

\begin{figure}[htbp]
\begin{center}
\includegraphics*[width=8cm]{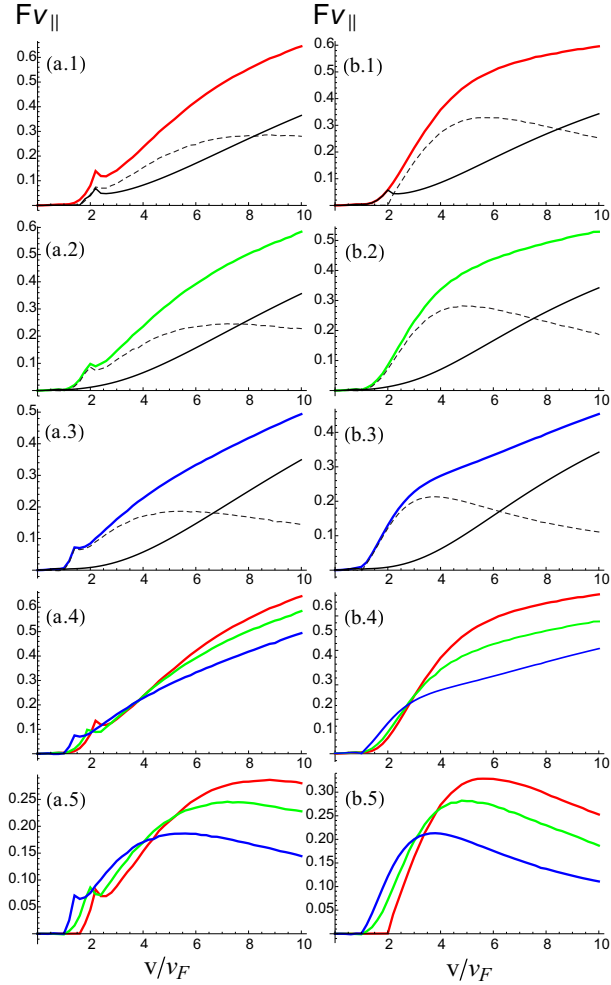}
\caption{(Color online)  Stopping power  in units of
$(Z e / 2 \pi)^2 \epsilon_s^{-1}$ as a function of charged particle velocity in
units of $v_F$. Panels (a) correspond to double layers, panels (b) to single  layer configuration.
Rows (1), (2) and (3) correspond to $E_g/\mu = \left\{ {0.0,1.0, 1.5}\right\}$,
 respectively, and are depicted by the red, green and blue curves, correspondingly.
The solid black curves are the particle-hole contributions, and the dashed curves
represent  plasmon contributions. Both (a.4) and (b.4) compare particle hole mode
contributions for the three values of the chosen energy gaps, and (a.5) and (b.5)
compare the corresponding plasmon contributions.}
\label{FIG:1}
\end{center}
\end{figure}

In Fig.~\ref{FIG:1}, we plot the rate of loss of energy  of a charged particle
 moving with speed $v$ at distance $z_0=2k_F^{-1}$  from a
 single and  double graphene layer configuration. 
For the pair of layers, we assume that
the energy gap is the same for both layers whose separation we chose as $d=k_F^{-1}$.
Both layers are equally doped to yield the chemical potential energy $\mu$, which is 
related to the Fermi level as $ \sqrt{\mu^2 +(E_g/2)^2}= \hbar v_F k_F$.
To evaluate the noninteracting polarization \eqref{eps}
all  frequencies in the polarization function are analytically continued
into the complex plane and acquire a small positive imaginary part
so that  $\omega \to \omega + i \gamma$. 
In that case the polarization assumes the form:
\begin{equation}
\label{EQ:NUMERICS}
\Pi (q,\omega+i \gamma) = i \pi
\frac{\Im \text{m}\Pi^{(0)}(q,\omega_p)}
{(\partial / \partial \omega)\Im \text{m}\Pi^{(0)}(q,\omega_p)\vert_{\omega= \omega_p}}
\delta\left({\omega- \omega_p}\right) + \Pi(q,\omega)
\end{equation}
where we have omitted the layer indices for brevity.
The plasmon resonances, denoted as  $\omega_p$, correspond to the poles of the inverse dielectric
function which are given by the solutions of $\epsilon(q,\omega_p - i \gamma)= 0$.
For infinitesimally small $\gamma = 0^+$ an analytical solution is readily available (See \cite{roslyak}
and references therein).
We chose  $\gamma/\mu = 10^{-4}$ to
account for possible lifetime broadening. 
The noninteracting polarization was calculated via adaptive local Monte-Carlo numerical integration.
The energy  loss  simulation is plotted in Figs.~\ref{FIG:1}
for various values of $E_g$.
\par

\begin{figure}[htbp]
\begin{center}
\includegraphics*[width=8cm]{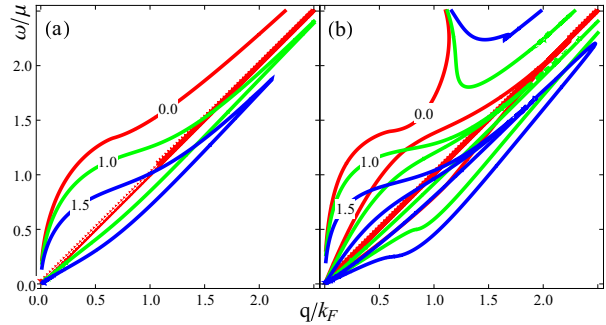}
\caption{(Color online) Undamped (two higher frequency branches) and damped (two lower in frequency branches)
plasmon dispersion relations for gapped graphene.  The curves in  (a)
correspond to  a single layer,  the curves in (b) to a double layer configuration. The red,
green and blue curves show the plasmon dispersion for $E_g/\mu = \left\{ {0.0,1.0, 1.5}\right\}$,
respectively.
 }
\label{FIG:2}
\end{center}
\end{figure}
 
The simulations shows that the plasmon
contribution to the energy loss is decreased with growing $E_g$, while the particle-hole 
contribution remains relatively unaffected by the gap as illustrated by Fig.\ \ref{FIG:1}. 
This may be attributed to
diminishing of the plasmon frequency $\omega_p(q) \sim \sqrt{q(1- (E_g/2\mu)^2)}$ 
as shown in Fig.\ \ref{FIG:2} (a).  
For the double layer configuration, a  small spike
at low velocity in the plasmon contribution to the energy loss is due to 
formation of the undamped
acoustic  (low frequency) and optical (high frequency)  plasmon branches.
The plasmon modes whose frequency is within the regions
\begin{gather}
\texttt{PH}_{>}= \hbar\omega > \sqrt{h^2 v^2_F q^2 + E^2_g}\\
\texttt{PH}_{<}= 0< \hbar \omega < \mu + \sqrt{\hbar^2 v_f^2 (q+k_F)^2 + (E_g/2)^2}
\end{gather} 
are Landau damped and given by $\Re\text{e} \epsilon(q,\omega_p - i \gamma) = 0$, and also depicted in Fig.~\ref{FIG:2}.
The damped plasmons contribute as particle-hole excitations.
\par
The plasmon contribution to the energy loss may be qualitatively explained by 
spanning the energy loss sector $0<\omega<q v$ through the plasmon dispersion if Fig.~\ref{FIG:1}. 
In the low velocity
regime, given by $\texttt{PH}_{<}(\omega  \rightarrow q v)$,
one has a contribution from the intraband particle-hole excitations which are relatively weak due to
the suppressed backscattering.
Once $v$ is increased so that the plasmons start to contribute one has rapid increase
in the lost energy. 
Once the sector fully engulfs the plasmon mode (modes) their contribution reaches its maximum
and becomes less than the particle-hole
mode contribution when $v/v_F\stackrel{>}{\sim}8$.
After that the increase in the energy loss is governed by the strong interband particle-hole excitations of $\texttt{PH}_{>}$.
Both types of the particle-hole excitations are given by the second term in Eq.~\eqref{EQ:NUMERICS}.
For gapless graphene the undamped plasmon may live only in the triangular region
where the distractive interference between the two types of the particle-hole excitation reduce the 
imaginary part of the polarization to zero.
The gap provides an additional space for the plasmons to run to the higher values of the wave vector.
In the double layer configuration       
both optical and acoustic plasmon modes contribute, but as $v$ is increased,
only the optical mode contributes. 
Since the spectral weight \footnote{the factor in front of the delta function in Eq.~\eqref{EQ:NUMERICS}.}
of the acoustical mode
is large compared with the optical branch, we see a spike once its contribution reaches the maximum.
\par
The effect of the energy gap on the plasmon contribution to the 
energy  loss  is summarized in Figs.\ \ref{FIG:2}(a.5) and  \ref{FIG:2}(b.5). 
These results show
intricate interplay between the plasmon and particle-hole contributions to the energy loss,
governed by the window in the particle-hole continuum is opened up in the presence of CPEF.
On one hand the gap reduces the velocity threshold of the collective plasma excitations.
The shorter wave length plasmon frequency become available in the window, and may be accommodated
by smaller energy loss sector. 
Furthermore, the plasmon branch decreases in frequency as $E_g$ is
increased, as seen in Fig.~\ref{FIG:2}, thereby making it more likely to excite
plasmons even when the charged particle speed is low,  and making gapped graphene to
have a higher stopping power than conventional gapless graphene. 
On the other hand, 
the plasmon contribution reaches its maximum for smaller velocities as and
the role these collective and particle-hole plasma excitations play in the energy loss are reversed in the
high-velocity limit. 
In this regime where the particle-hole excitations should dominate the window 
makes the stopping power of gapped graphene being less than when there is no gap.

\section*{Acknowledgement(s)}

This research was supported by  contract \# FA 9453-07-C-0207 of AFRL. Dr. Huang would like
 to thank the Air Force Office of Scientific Research
(AFOSR) for its support.

\newpage

\appendix

\section{Appendix A}

\begin{table}[htdp]
\caption{Free vs dressed Dirac electron states.}
\begin{center}
\begin{tabular}{|c|c|c|}
\hline
 & \large Free Dirac & \large Dressed Dirac\\
\hline
Energy &
$
E_{\mathbf{k}} = \pm k
$
& $E_{\mathbf{k}} = N_0 \omega_0 \pm \sqrt{(E_g/2)^2 + k^2}$\\
\hline
Wave &
$
\left({
\begin{array}{c}
\psi^{+}_{\mathbf{k}}\\
\psi^{-}_{\mathbf{k}}
\end{array}
}\right) = \frac{\texttt{e}^{i \mathbf{kr}}}{\sqrt{2}}
\left({
\begin{array}{cc}
1 & \texttt{e}^{i \theta_{\mathbf{k}}}\\
1 & -\texttt{e}^{i \theta_{\mathbf{k}}}
\end{array}
}\right)
\left({
\begin{array}{c}
\vert{\uparrow}\rangle\\
\vert{\downarrow}\rangle
\end{array}
}\right)
$
&
$
\left({
\begin{array}{c}
\psi^{+}_{\mathbf{k}}\\
\psi^{-}_{\mathbf{k}}
\end{array}
}\right) = \frac{\texttt{e}^{i \mathbf{kr}}}{\sqrt{1 + \alpha^2_{N0,k}}}
\left({
\begin{array}{cc}
1 & \alpha_{N0,k}\texttt{e}^{i \theta_{\mathbf{k}}}\\
\alpha_{N0,k} & -\texttt{e}^{i \theta_{\mathbf{k}}}
\end{array}
}\right)
\left({
\begin{array}{c}
\vert{+\frac{1}{2},N_0}\rangle\\
\vert{-\frac{1}{2},N_0}\rangle
\end{array}
}\right)
$\\
Function & & \\
\hline
Structure &
$
\mathcal{F}^{\beta}_{\mathbf{k},\mathbf{k+q}} =
\frac{1}{2} (1 \pm \cos{\theta_{k,k+q}})
$
&
$
\begin{array}{c}
\mathcal{F}^{\beta}_{\mathbf{k},\mathbf{k+q}} =
\frac{1}{(1+\alpha^2_{N0,k})(1+\alpha^2_{N0,k+q})}
(1 + \alpha^2_{N0,k} \alpha^2_{N0,k+q}\pm\\
\pm 2 \alpha_{N0,k} \alpha_{N0,k+q}\cos{\theta_{k,k+q}}
\end{array}
$\\
Factor & &\\
\hline
Distribution &
$
f^{\beta}_{k} = \frac{\exp{-\left({E_{\mathbf{k}}-\mu}\right)/k_B T}}{1+\exp{-\left({E_{\mathbf{k}}-\mu}\right)/k_B T}}
$
&
$
f^{\beta}_{k} = \frac{\exp{-\left({E_{\mathbf{k}}-\mu-N_0 \omega_0}\right)/k_B T}}{1+\exp{-\left({E_{\mathbf{k}}-\mu-N_0 \omega_0}\right)/k_B T}}
$\\
& & \\
\hline
Parameters &
 $
 \theta_{k} = \tan^{-1} {\left({\frac{k_x}{k_y}}\right)}
 $
 &
 $
 \alpha_{N0,k} = \frac{k}{\sqrt{k^2+(E_g/2)^2}+E_g/2}
 $\\
 &
 $
 \vert{\uparrow}\rangle =
 \left({
 \begin{array}{c}
 1\\
 0
 \end{array}
 }\right);
 \vert{\downarrow}\rangle =
 \left({
 \begin{array}{c}
 0\\
 1
 \end{array}
 }\right)
 $
 &
 $
\left({
\begin{array}{c}
\vert{+\frac{1}{2}N_0}\rangle\\
\vert{-\frac{1}{2}N_0}\rangle
\end{array}
}\right) =
\left({
\begin{array}{cc}
\cos{\left({\frac{\Phi}{2}}\right)} & \sin{\left({\frac{\Phi}{2}}\right)}\\
-\sin{\left({\frac{\Phi}{2}}\right)} & \cos{\left({\frac{\Phi}{2}}\right)}
\end{array}
}\right)
\left({
\begin{array}{c}
\vert{\uparrow,N_0}\rangle\\
\vert{\downarrow,N_0}\rangle
\end{array}
}\right)
$\\
& &\\
&
$
\cos{\theta_{k,k+q}} = \frac{k+q \cos{\theta_{k,q}}}{\sqrt{k^2+q^2+2 k q \cos{\theta_{k,q}}}}
$
&
$
\tan^2{\left({\frac{\Phi}{2}}\right)} = \frac{\sqrt{\omega^2_0 +4 W_0^2 (N_0+1)/N_0}-\omega_0}{\sqrt{\omega^2_0 +4 W_0^2 (N_0+1)/N_0}+\omega_0}
$
\\
\hline
\end{tabular}
\end{center}
\label{default}
\end{table}%

\end{document}